\documentclass[%
 reprint, 
superscriptaddress,
 amsmath,amssymb,
 aps,
prb,
floatfix,
]{revtex4-1}

\usepackage{graphicx} 
\usepackage{amsmath}
\usepackage{bm}       
\usepackage{placeins} 
\usepackage[separate-uncertainty=true, multi-part-units=single]{siunitx}

\graphicspath{{figures/rendered/}}

\renewcommand{\vec}[1]{\boldsymbol{\mathrm{#1}}}

\begin{document}


\title{Time- and momentum-resolved phonon population dynamics \\with ultrafast electron diffuse scattering}

\author{Laurent P. Ren\'{e} de Cotret}
\author{Jan-Hendrik P\"{o}hls}
\author{Mark J. Stern}
\author{Martin R. Otto}
\author{Mark Sutton}
\affiliation{%
 Department of Physics, Center for the Physics of Material,\\
 McGill University, 3600 rue University, Montr\'{e}al, QC, CA 
}
\author{Bradley J. Siwick}
\email{bradley.siwick@mcgill.ca}
\affiliation{%
 Department of Physics, Center for the Physics of Material,\\
 McGill University, 3600 rue University, Montr\'{e}al, QC, CA 
}
\affiliation{%
Department of Chemistry, McGill University, \\
801 rue Sherbrooke Ouest, Montr\'{e}al, QC, CA
}

\date{\today}

\begin{abstract}
Interactions between the lattice and charge carriers can drive the formation of phases and ordering phenomena that give rise to conventional superconductivity, insulator-to-metal transitions, and charge-density waves. These couplings also play a determining role in properties that include electric and thermal conductivity. Ultrafast electron diffuse scattering (UEDS) has recently become a viable laboratory-scale tool to track energy flow into and within the lattice system across the entire Brillouin zone, and to deconvolve interactions in the time domain. Here, we present a detailed quantitative framework for the interpretation of UEDS signals, ultimately extracting the phonon mode occupancies across the entire Brillouin zone. These transient populations are then used to extract momentum- and mode-dependent electron-phonon and phonon-phonon coupling constants. Results of this analysis are presented for graphite, which provides complete information on the phonon-branch occupations and a determination of the $A_1'$ phonon mode-projected electron-phonon coupling strength $\langle g_{e,A_1'}^2 \rangle = \SI{0.035 \pm 0.001}{\square\electronvolt}$ that is in agreement with other measurement techniques and simulations.
\end{abstract}
\maketitle

\section{Introduction}
Elementary excitations and their mutual couplings form the fundamental basis of our understanding of diverse phenomena in condensed matter systems. The interactions between collective excitations of the lattice system (phonons) and charge carriers, specifically, are known to lead to superconductivity, charge-density waves, multiferroicity, and soft-mode phase transitions~\cite{eliashberg1960interactions, hur2004electric}. These carrier-phonon interactions are also central to our understanding of electrical transport, heat transport, and energy conversion processes in photovoltaics and thermoelectrics~\cite{Zhao2016}. Phonons can themselves be intimately mixed in to the very nature of more complex elementary excitations, as they are in polarons or polaritons, or intertwined with electronic, spin, or orbital degrees-of-freedom, as it now seems is the case for the emergent phases of many strongly-correlated systems that exhibit complex phase diagrams like high-$T_c$ superconductors~\cite{Miyata2017,kim2012ultrafast, lanzara2001evidence}.

Our inability to fully characterize the nature of elementary excitations and to quantify the strength of their momentum-dependent interactions has been one of the primary barriers to our understanding of these phenomena, particularly in complex anisotropic materials. Ultrafast pump-probe techniques provide an opportunity to study couplings between elementary excitations rather directly. Photoexcitation can prepare a non-equilibrium distribution of quasiparticles or other selected modes whose subsequent relaxation dynamics and coupling to other degrees of freedom can be followed in the time-domain. Under favourable circumstances, qualitatively distinct channels can be disentangled by their associated spectra (response functions) and time-scales. This field has evolved rapidly over the last decade, both from the perspective of the selectivity of the initial excitation and the ability to probe the subsequent dynamics over a broad range of frequencies. For example, spectroscopic pump-probe techniques in the terahertz range have been used to interrogate the link between electrons/holes and optical phonons in hybrid lead halide perovskites~\cite{Lan2019} and to investigate the time-ordering of phenomena behind charge-density waves in titanium diselenide~\cite{Porer2014}.

The low-photon momentum associated with optical frequencies, however, prevents the most commonly applied optical photon-in, optical photon-out techniques from providing a full characterization of the wavevector-dependent interactions between elementary excitation. Time-resolved Raman and Brillouin spectroscopies, for example, are limited to the interrogation of zero-momentum (zone-center) phonons for this reason~\cite{Tsen2009, Yan2009, Yang2017}.

In recent years, non-optical ultrafast techniques have been developed to probe wavevector-dependent dynamics. The most mature of these approaches is time- and angle-resolved photoemission spectroscopy (trARPES), which has been used to assemble a complete picture of the dynamics of the electronic and spin excitations following the photoexcitation of materials \cite{Johannsen2013, Gierz2015, Stange2015, Yang2017, Rohde2018}.

The ability to directly interrogate wavevector-dependent dynamics within the phonon system, on the other hand, is an extremely recent development. Ultrafast X-ray diffuse scattering \cite{Trigo2010, Zhu2015, Wall2018} is one technique that has the potential to reveal lattice excitation dynamics across the whole Brillouin zone. This approach leverages the remarkable brightness of the beams available from the current generation of X-ray free-electron laser facilities to measure the time dependence of the diffuse (phonon) scattering from materials following photoexcitation.

At the laboratory scale, there has been similar progress made in furthering ultrafast electron beam brightness which has enabled equivalent diffuse scattering experiments to be performed. Ultrafast electron diffuse scattering (UEDS) has the potential to be transformative in that it can provide both a wavevector-resolved view of the coupling between electron and lattice systems \cite{Harb2016, Chase2016, Waldecker2017, Stern2018, Konstantinova2018} and the wavevector dependence of the interactions within the phonon system itself. The large scattering cross-section of electrons, combined with the relative flatness of the Ewald sphere, potentially allows for the simultaneous measurement of both the average lattice structure (via Bragg scattering) and lattice excitation dynamics (via diffuse scattering) in specimens as thin as a single atomic layer. 

In this work we provide a description of the signals contained in UEDS measurements and a comprehensive and broadly applicable computational method for UEDS data reduction based on density functional perturbation theory (DFPT). Specifically, we present a procedure to recover phonon population dynamics as a function of the phonon branch and wavevector, and a determination of wavevector-dependent (or mode-projected) electron-phonon coupling constants from those phonon population measurements. This method uses only the measured time-resolved UEDS patterns and DFPT determinations of the phonon polarization vectors as inputs. The application of this approach to the case of photodoped carriers in the Dirac cones of thin graphite is demonstrated. The electron-phonon coupling strength to the strongly-coupled $A_1'$ phonon at the $\vec{K}$-point of the Brillouin zone, and the nonequilibrium optical and acoustic phonon branch populations as a function of time following excitation across the whole Brillouin zone are all determined from the UEDS measurements.
\section{Experimental and Computational methods}

\begin{figure*}
	\centering
	\includegraphics[width=1\textwidth]{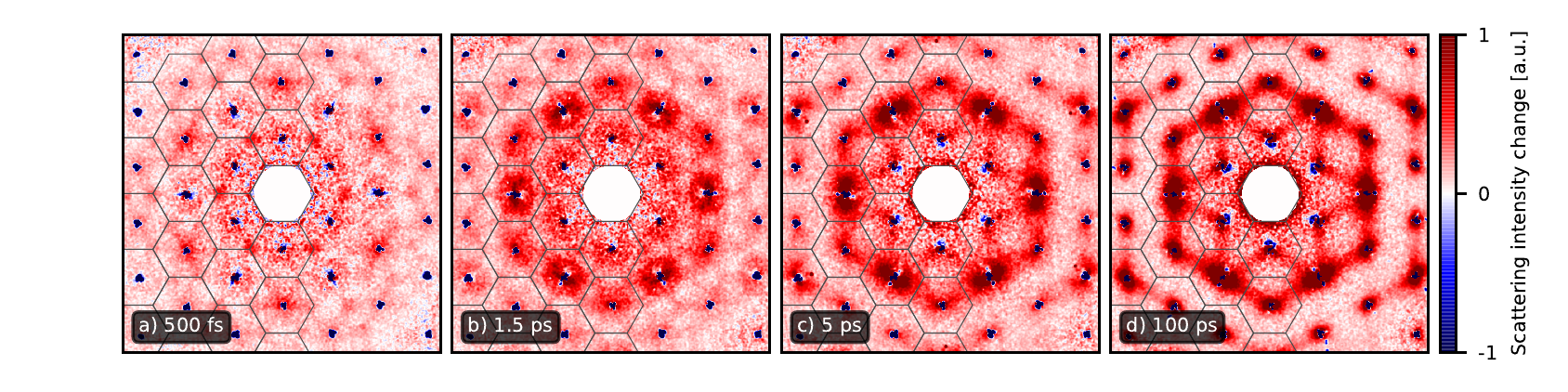}
	\caption{Experimental change in transient electron scattering intensity $\Delta I(\vec{q}, \tau) = I(\vec{q},t_0+\tau) - I(\vec{q}, t_0)$ of photoexcited graphite for a few representative time-delays. This figure shows the increase (red) in diffuse scattering intensity across a wide range of reciprocal space ($|\vec{q}|<\SI{12}{\per\angstrom}$). Negative changes (blue) are limited to Bragg peaks. Brillouin zones have been overlaid on half of the detector to highlight diffuse intensity structure. Scattering intensities are rotationally averaged as described in Section \ref{SEC:corrections}.}	
	\label{FIG:graphite}
\end{figure*}

The change in experimental electron scattering intensity of graphite, photoexcited with \SI{35}{\femto\second} pulses of \SI{800}{\nano\meter} light at a fluence of \SI{12}{\milli\J\per\square\centi\meter}, are presented in Figure \ref{FIG:graphite} for a few representative time-delays. This section provides details on the experimental parameters, data processing steps, and computational techniques that are used in this work.
\begin{figure}
	\centering
	\includegraphics[width=3.38in]{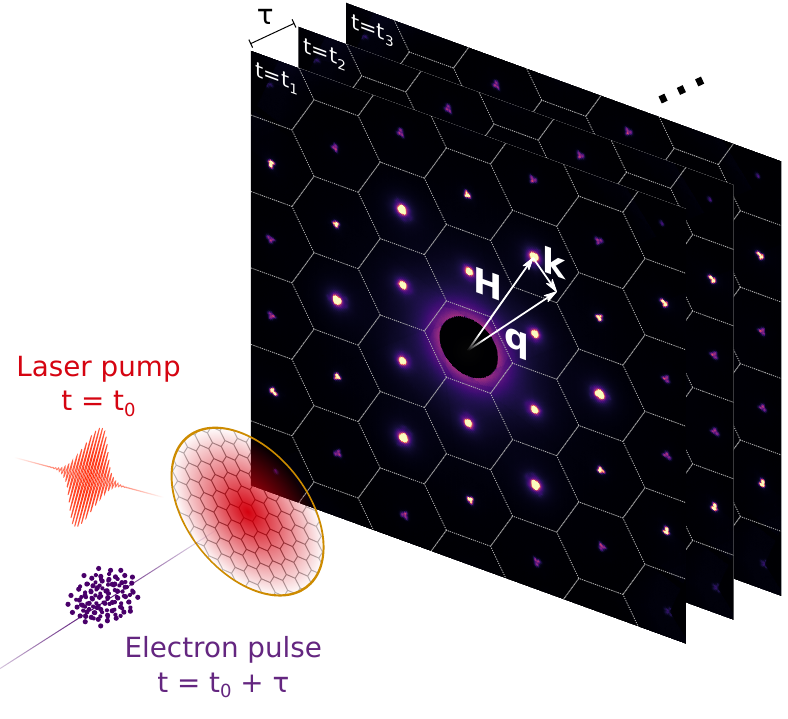}
	\caption{Schematic diagram of ultrafast electron diffuse scattering experiments. Samples are photoexcited with ultrashort pulses of light at time $t=t_0$. After some delay $\tau$, an ultrashort electron pulse scatters through the sample. Elastic and inelastic signals are collected in transmission geometry. By scanning time-delays $\tau$, a stroboscopic movie of the dynamics can be assembled. This figure also shows the example of a scattering vector $\vec{q}$, and its associated reduced wavevector $\vec{k}$, which are related by the nearest Bragg reflection $\vec{H}$.}
	\label{FIG:intro}
\end{figure}
\subsection{Data acquisition}
UEDS measurements are pump-probe experiments in which an ultrafast laser pulse is used to photoexcite a thin single-crystal specimen at $t=t_0$, followed by probing the specimen with an ultrafast electron pulse at $t=t_0 + \tau$, resulting in the acquisition of a transmission electron scattering pattern. By scanning across time-delays $\tau$, the dynamics in the \si{\femto\second} -- \si{\nano\second} ($10^{-15} \textrm{--} 10^{-9}$ \si{\second}) range can be recorded. UEDS data can be acquired coincidentally during ultrafast electron diffraction (UED) experiments with state-of-the-art detection cameras, although UEDS intensities are empirically $10^4$ to $10^6$ times less intense than those of Bragg diffraction. UEDS measurements are inherently statistical. Pump-probe experiments sample many decay processes, incoherently in time. Hence, all possible decay channels are represented, proportionally to their statistical likelihood.

Scattering measurements presented in this work use bunches of $10^7$ electrons, accelerated to \SI{90}{\kilo\electronvolt}, at a repetition rate of \SI{1}{\kilo\hertz}. A radio-frequency cavity is used to compress electron bunches to $\approx \SI{150}{\femto\second}$ at the sample, as measured by a home-built photoactivated streak camera~\cite{kassier2010compact}. More detailed descriptions of this instrument are given elsewhere~\cite{Chatelain2012, Morrison2013, Otto2017}. Analysis of static diffraction patterns indicate a momentum resolution of \SI{0.06}{\per\angstrom}, while the range of visible reflections is consistent with a real-space resolution of $<\SI{1}{\angstrom}$. \SI{35}{\femto\second} pump laser pulses of \SI{800}{\nano\meter} (\SI{1.55}{\electronvolt}) light are used to photoexcite a single-crystal flake of freestanding single crystal natural graphite, provided by Naturally Graphite. The flakes were mechanically exfoliated to a thickness of \SI{70}{\nano\meter}. 

The interrogated film area covers \SI{500 x 500}{\micro\meter}, with a pump spot of \SI{1 x 1}{\milli \meter} full-width half-max (FWHM) ensuring nearly uniform illumination of the probed volume. The film was pumped at a fluence of \SI{12}{\milli\J\per\square\centi\meter}, resulting in an absorbed energy density of \SI{8}{\J\per\cubic\meter}. The scattering patterns are collected with a Gatan Ultrascan 895 camera; a \SI{2.54 x 2.54}{\centi\meter} phosphor screen fiber-coupled to a \SI{2048 x 2048}{px} charge-coupled detector (CCD) placed \SI{25}{\centi\meter} away from the sample.
The experiment herein consists of time-delays in the range of \SIrange{-40}{680}{\pico\second}. Per-pixel scattering intensity fluctuations over laboratory time reveals a transient dynamic range of $1:10^8$, allowing the acquisition of diffraction patterns and diffuse scattering patterns simultaneously~\footnote{The intensity fluctuations of pixel values across scattering patterns acquired before photoexcitation are $10^8$ times smaller than the brightest Bragg reflection}. 
\subsection{Processing and corrections}
\label{SEC:corrections}
Scattering intensity patterns are inherently redundant due to the point-group symmetry of the scattering crystal. When this symmetry is not broken by photoexcitation or the dynamical phenomena itself, it is possible to use this redundancy to enhance the signal to noise ratio of a UEDS data set. No dynamical phenomena breaking point-group symmetry was observable within the raw signal to noise of the current measurements, so the measured patterns have been subject to a six-fold discrete azimuthal average based on the $D_{6h}$ point-group of graphite. This discrete rotational average effectively increases the signal-to-noise ratio of this data set by a factor of $\sqrt{6}$ and is therefore an effective data processing step given the weakness of the diffuse scattering signals. ``Scattering intensities" is henceforth implied to mean six-fold averaged scattering intensities. 

Scattering from a few samples with varying thicknesses (\SIrange{10}{90}{\nano\meter}) was acquired. There was no quantitative difference in the Bragg scattering dynamics, indicating that scattering from these samples is kinematical to a good approximation. The expected effects of multiple scattering on the diffuse scattering intensity distribution will be discussed further below.
\subsection{Computational methods}
Structure relaxation was performed using the plane-wave self-consistent field program \texttt{PWscf} from the Quantum ESPRESSO software suite~\cite{Giannozzi2017}. The graphite structure was fully relaxed using a \num{18 x 18 x 10} $\vec{k}$-point mesh centered at $\vec{\Gamma}$ and force (energy) threshold of \SI{1e-8}{Ry/Bohr} (\SI{1e-14}{Ry}). The dynamical matrices were computed on \num{5 x 3 x 3} $\vec{q}$-point grid using a self-consistency threshold of \SI{1e-18}{Ry}. The resulting graphite structure has the following lattice vectors:
\begin{equation*}
	\vec{a}_1 = a ~ \vec{e}_1, ~~~~
	\vec{a}_2 = \frac{a}{2} \left( \sqrt{3} ~ \vec{e}_2 - \vec{e}_1 \right), ~~~~
	\vec{a}_3 = c ~ \vec{e}_3,
\end{equation*}
where $a=\SI{2.462}{\angstrom}$, $c=\SI{6.837}{\angstrom}$, and $\left\{ \vec{e}_i \right\}$ are the usual Euclidean vectors. Graphite has four atoms in the unit cell, with two groups of two atoms forming lattices rotated with respect to each other. This structure respects the symmetries of the $D_{6h}$ point group~\footnote{The space group of this structure is $P6_3/mmc$ (Hermann-Mauguin symbol) or $D_{6h}^4$ (Schoenflies symbol).}.

The phonon frequencies $\left\{ \omega_{j,\vec{k}} \right\}$ and polarization vectors $\left\{ \vec{e}_{j,s,\vec{k}} \right\}$ were computed using the \texttt{PHonon} program in the Quantum ESPRESSO software suite, using the B86b exchange-coupled Perdew-Burke-Ernzerhof (B86bPBE) generalized gradient approximation (GGA) and the projector augmented-wave (PAW) method~\cite{Becke1986, Perdew1996, Blochl1994}. The cutoff-energy of the wavefunction was set to \SI{100}{Ry}, while the cutoff energy for the charge density was set to \SI{1200}{Ry}, and a Fermi-Dirac smearing of \SI{0.06}{Ry} was applied. To include the dispersion energy between the carbon layers, the exchange-hole dipole moment (XDM) method was used~\cite{Becke2007}.
\section{Theory}
Similar to X-ray scattering, under the kinematical approximation the measurement of the total scattering intensity at scattering vector $\vec{q}$ and time $t$, $I(\vec{q}, t)$, of an electron bunch interacting with a thin film of crystalline material, can be separated as follows:
\begin{equation*}
I(\vec{q}, t) = I_0(\vec{q}, t) + I_1(\vec{q}, t) + ... ~ ,
\end{equation*}
where the intensity $I_n$ represents the scattered intensity of an electron that interacted with $n$ phonons. Specifically, $I_0$ represents diffraction, or Bragg scattering, and $I_1$ represents the first-order \emph{diffuse scattering intensity}. The experimentally-observed ratio $I_0/I_1$ ranges between $10^4$ -- $10^6$~\footnote{Detector counts for the brightest Bragg peak reaches as much as 20 000 counts, while the average diffuse feature shown in Figure \ref{FIG:graphite} is 0.2 counts.}. Higher-order terms have much smaller cross-sections, hence much lower contribution to scattering intensity, and are therefore ignored in this work. The expressions for the intensities $I_0$ and $I_1$ are given below:
\begin{align}
I_0(\vec{q}, t) &= N_c I_e \left| \sum_s f_s(\vec{q})~e^{-W_s(\vec{q}, t)}~e^{-i [\vec{q} \cdot \vec{R}_s(t)]} \right|^2 \\
I_1(\vec{q}, t) &= N_c I_e \sum_j \frac{n_{j,\vec{k}}(t) + 1/2}{\omega_{j,\vec{k}}(t)} |F_{1j}(\vec{q},t)|^2 \label{EQ:diffuse}
\end{align}
where $N_c$ is the number of diffracting cells, $I_e$ is the intensity of scattering from a single event, $\vec{q}$ is the wavevector (or scattering vector), $\vec{k} = \vec{q} - \vec{H}$ is the reduced wavevector associated to $\vec{q}$ with respect to the nearest Bragg reflection $\vec{H}$ (see Figure \ref{FIG:intro}), $s$ are indices associated with atoms in the crystal unit cell, $\vec{R}_s(t)$ is the real-space atomic position of atom $s$, $W_s(\vec{q}, t)$ is the Debye-Waller factor of atom $s$, $f_s(\vec{q})$ are the atomic form factors, $j \in \{1, 2, ..., N \}$ runs over phonon modes, and $\left\{ n_{j,\vec{k}}(t) \right\}$ and $\left\{ \omega_{j,\vec{k}}(t) \right\}$ are the population and frequency associated with phonon mode $j$, respectively~\cite{Wang1995, Xu2005}.

The diffuse scattering intensity contribution of each mode $j$ is weighted by a factor, called the \emph{one-phonon structure factor} $|F_{1j}(\vec{q},t)|^2$:
\begin{equation}
|F_{1j}(\vec{q}, t)|^2 = \left|\sum_s e^{-W_s(\vec{q}, t)} \frac{f_s(\vec{q})}{\sqrt{\mu_s}} (\vec{q} \cdot \vec{e}_{j,s,\vec{k}})\right|^2
\label{EQ:oneph}
\end{equation}
where $\mu_s$ is the mass of atom $s$, and $\left\{ \vec{e}_{j, s,\vec{k}} \right\}$ are the wavevector-dependent polarization vectors associated with phonon mode $j$ for atom $s$. The one-phonon structure factors represent the contribution of phonon mode $j$ on the overall intensity at a specific scattering vector $\vec{q}$ and time $t$. $|F_{1j}(\vec{q}, t)|^2$ are a measure of two things: the locations in Brillouin zone where the phonon mode polarization vectors $\left\{ \vec{e}_{j, s,\vec{k}} \right\}$ are aligned in such a way that they will contribute to diffuse scattering intensity on the detector, expressed via the terms $\left\{ \vec{q} \cdot \vec{e}_{j,s,\vec{k}} \right\}$; and the strength of the contribution of a single scattering event, including the effect of the instantaneous disorder in the material, expressed via the quantities $\left\{ e^{-W_s(\vec{q}, t)} f_s(\vec{q})/\sqrt{\mu_s} \right\}$. 

Instantaneous disorder is described by the Debye-Waller factors $\left\{ W_s(\vec{q}, t) \right\}$. The general expression of the \emph{anisotropic} Debye-Waller factor for atom $s$, $W_s(\vec{q}, t)$, is given below:
\begin{equation}
W_s(\vec{q}, t) = \frac{1}{4\mu_s} \sum_{j, \vec{k}} |a_{j, \vec{k}}(t)|^2 ~|\vec{q} \cdot \vec{e}_{j,s,\vec{k}}|^2
\label{EQ:Debye_Waller}	
\end{equation}
where $a_{j, \vec{k}}(t)$ is the phonon mode vibration amplitude for mode $j$ at reduced wavevector $\vec{k}$~\cite{Xu2005}:
\begin{equation}
|a_{j, \vec{k}}(t)|^2 = \frac{2 \hbar}{N_c \omega_{j, \vec{k}}(t)}  \left(n_{j,\vec{k}}(t) + \frac{1}{2}\right).
\end{equation}
Debye-Waller factors describe the reduction of intensity at scattering vector $\vec{q}$ due to the effective deformation of a single atom's scattering potential that results from the collective lattice vibrations in \emph{all} phonon modes. Wavevector-specific information is in general impossible to extract from the transient changes to the Debye-Waller factors that result from photoexcitation and the non-equlibrium phonon populations that such excitation produces.

The expression for $I_1(\vec{q},t)$ in Equation \eqref{EQ:diffuse} (and related quantities) apply rigorously under single electron scattering conditions. The most probable type of multiple scattering event affecting $I_1(\vec{q}, t)$ is diffuse scattering followed by secondary Bragg scattering, not multiple or consecutive diffuse scattering events~\cite{Cowley1979, Wang1995}. A secondary Bragg scattering event only changes the electron wavevector by a reciprocal lattice vector; thus, this type of multiple scattering results in a redistribution of diffuse intensity from lower-order to higher-order Brillouin zones (further from $|\vec{q}|=0$). However, the wavevector dependence of experimental diffuse intensities is not strongly influenced, even under experimental conditions where such dynamical effects are important. The strength of the scattering selection rules implied by the $\left\{ \vec{q} \cdot \vec{e}_{j,s,\vec{k}} \right\}$ terms, and described further below, are reduced as the proportion of multiple scattering increases.
\section{Results and Discussion}
In this section a comprehensive procedure for ultrafast electron diffuse scattering data reduction will be presented. This approach recovers the time- and wavevector-dependent phonon population dynamics in each of the phonon branches. In addition, the population dynamics of the $A_1'$ phonon, a strongly-coupled optical phonon in graphite, is used to demonstrate the extraction of wavevector-dependent (mode-projected) electron-phonon coupling constants from ultrafast electron diffuse intensities.
\subsection{Calculation of phonon polarization vectors and eigenfrequencies}
The quantitative connection between the observed diffuse intensity $I_1(\vec{q},t)$ and the phonon populations, $\left\{ n_{j,\vec{k}} \right\}$, is provided by the one-phonon structure factors, $|F_{1j}(\vec{q}, t)|^2$. A determination of $|F_{1j}(\vec{q}, t)|^2$ requires phonon polarization vectors $\left\{ \vec{e}_{j,s,\vec{k}} \right\}$ and associated frequencies $\left\{ \omega_{j,\vec{k}} \right\}$. Density functional perturbation theory (DFPT) is a widely used, readily-available method to compute these phonon properties. 

The separation of frequencies and polarization vectors into modes is key to the calculation of one-phonon structure factors $|F_{1j}(\vec{q},t)|^2$. The phonon frequencies $\left\{ \omega_{j,\vec{k}} \right\}$ and polarization vectors $\left\{ \vec{e}_{j,s,\vec{k}} \right\}$ were computed independently at every $\vec{k}$; however, diagonalization routines have no way of clustering eigenvalues and eigenvectors into physically-relevant groups (i.e. phonon branches). Association between atomic motions (given by polarization vectors) and a particular mode is only well-defined near the $\vec{\Gamma}$-point~\cite{Paulatto2013}. Clustering of phonon properties into phonon branches is described in detail in Appendix \ref{AP:mode_clustering}.

The polarization vectors and frequencies, calculated for irreducible $\vec{k}$-points~\footnote{The coverage of irreducible $\vec{k}$-points is important. Only computing phonon properties along high-symmetry lines is fraught with peril, given that polarization vectors can vary significantly not only along high-symmetry lines, but over the entire Brillouin zone.}, were extended over the entire reciprocal space based on crystal symmetries using the \texttt{crystals} software package~\cite{RenedeCotret2018}.
\subsection{Debye-Waller calculation}
\label{SEC:debye_waller}
A key component of the computation of one-phonon structure factors $|F_{1j}(\vec{q}, t)|^2$ is the calculation of the Debye-Waller factors $\left\{ W_s(\vec{q}, t) \right\}$, representing the instantaneous disorder of the material. Based on Equation \eqref{EQ:Debye_Waller}, terms of the form $\left\{ W_s(\vec{q},t) \right\}$ are not sensitive reporters on the wavevector dependence of nonequilibrium phonon distributions because their value at every $\vec{q}$ involves a sum of all mode, at every reduced wavevector $\vec{k}$. Phonon population dynamics can only affect the magnitude of the Debye-Waller factors. The potential time dependence of the Debye-Waller factors was investigated, via the time dependence of mode populations $\left\{ n_{j,\vec{k}} \right\}$. Profoundly non-equilibrium distributions of phonon modes were simulated, with all modes populated equivalently to a temperature of \SI{300}{\kelvin} except one mode at high temperature~\footnote{A maximum of \SI{5000}{\kelvin} for optical modes, and \SI{1000}{\kelvin} for acoustic modes. The discrepancy between maximum temperatures represents the fact that the heat capacity of acoustic modes is much higher.}. These extreme non-equilibrium distributions increased the value the terms $\sum_s W_s(\vec{q},t)$ by at most \SI{1.5}{\percent} for optical modes, and \SI{5}{\percent} for acoustic modes. Since these fractional changes are constant across $\vec{q}$, wavevector-dependent changes in UEDS signals are not impacted significantly by transient changes to the Debye-Waller factors and any time dependence of the one-phonon structure factors $|F_{1j}(\vec{q},t)|^2$ that result from the Debye-Waller factors themselves can be ignored to a good approximation.
\subsection{One-phonon structure factors}
\begin{figure}
	\centering
	\includegraphics[width=1\columnwidth]{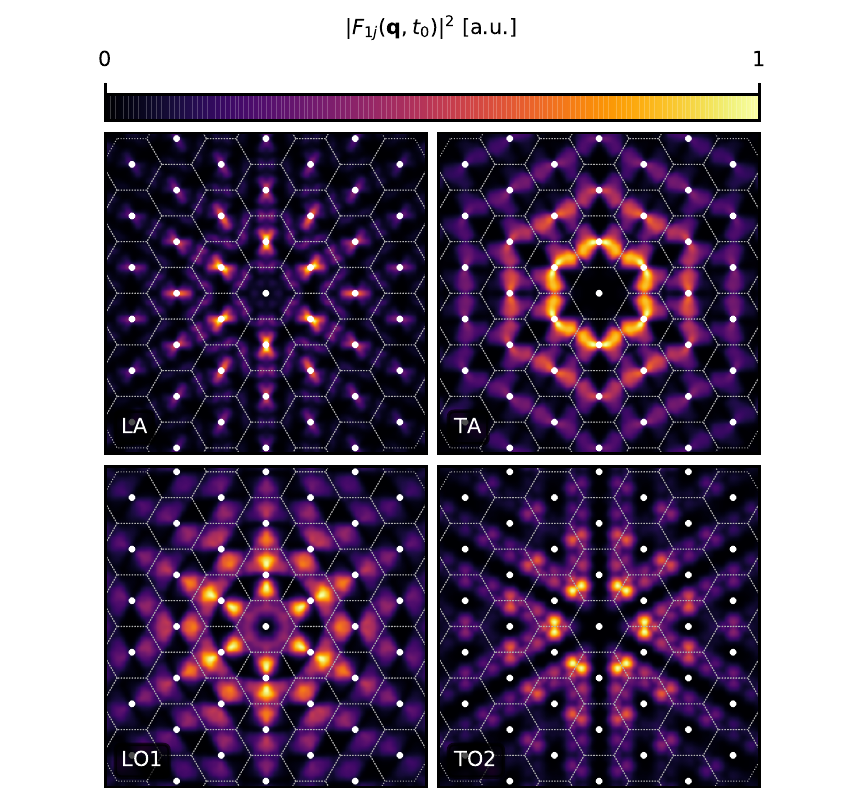}
	\caption{Calculated one-phonon structure factors $|F_{1j}(\vec{q}, t)|^2$ of selected in-plane modes of graphite, at \SI{300}{\kelvin} ($t=t_0$), for $\vec{q}$ vectors equivalent to the detector area shown in Figure \ref{FIG:graphite}. Bright spots indicate locations in reciprocal space where the associated mode contributes strongly to the diffuse scattering intensity.Brillouin zone outlines are overlaid, and their centers (Bragg peaks) are marked with a white dot. While one would expect a wavevector-dependent behavior $|F_{1j}(\vec{q},t)|^2 \propto |\vec{q}|^2$ from Equation \eqref{EQ:oneph}, the gaussian nature of the Debye-Waller factor terms $\left\{ W_s(\vec{q},t) \right\}$ and of the atomic form factor terms $\left\{ f_s(\vec{q})\right\}$ decrease the amplitude of the one-phonon structure factors at larger $\vec{q}$.}
	\label{FIG:phonon_struct_factor}
\end{figure}

Using the computed Debye-Waller factors from the previous section, the calculation of $|F_{1j}(\vec{q}, t)|^2$ was carried out, from Equation \eqref{EQ:oneph}, for the eight in-plane phonon modes of graphite: the longitudinal modes LA, LO1 -- LO3, and the transverse modes TA, TO1 -- TO3~\footnote{The calculation of $|F_{1j}(\vec{q}, t)|$ is trivial for out-of-plane modes ZA, ZO1 -- ZO3  because $\vec{q} \cdot \vec{e}_{j,s,\vec{k}} \equiv 0$ for these modes.}. The resulting one-phonon structure factors of a few in-plane modes, with occupations equivalent to a temperature of \SI{300}{\kelvin}, are shown in Figure \ref{FIG:phonon_struct_factor}. One-phonon structure factors $|F_{1j}(\vec{q},t)|^2$ display striking scattering vector dependence (selection rules) based on the nature of phonon polarization vectors $\left\{ \vec{e}_{j,s,\vec{k}} \right\}$. Specifically, near $\vec{\Gamma}$, the one-phonon structure factor for longitudinal modes is highest in the radial direction, because the polarization of those modes is parallel to $\vec{q}$. On the other hand, $|F_{1j}(\vec{q},t)|^2$ for transverse modes is highest (near $\vec{\Gamma}$) in the azimuthal direction for transverse modes, because the polarization of those modes is perpendicular to $\vec{q}$.

\begin{figure*}
	\centering
	\includegraphics[width=1\textwidth]{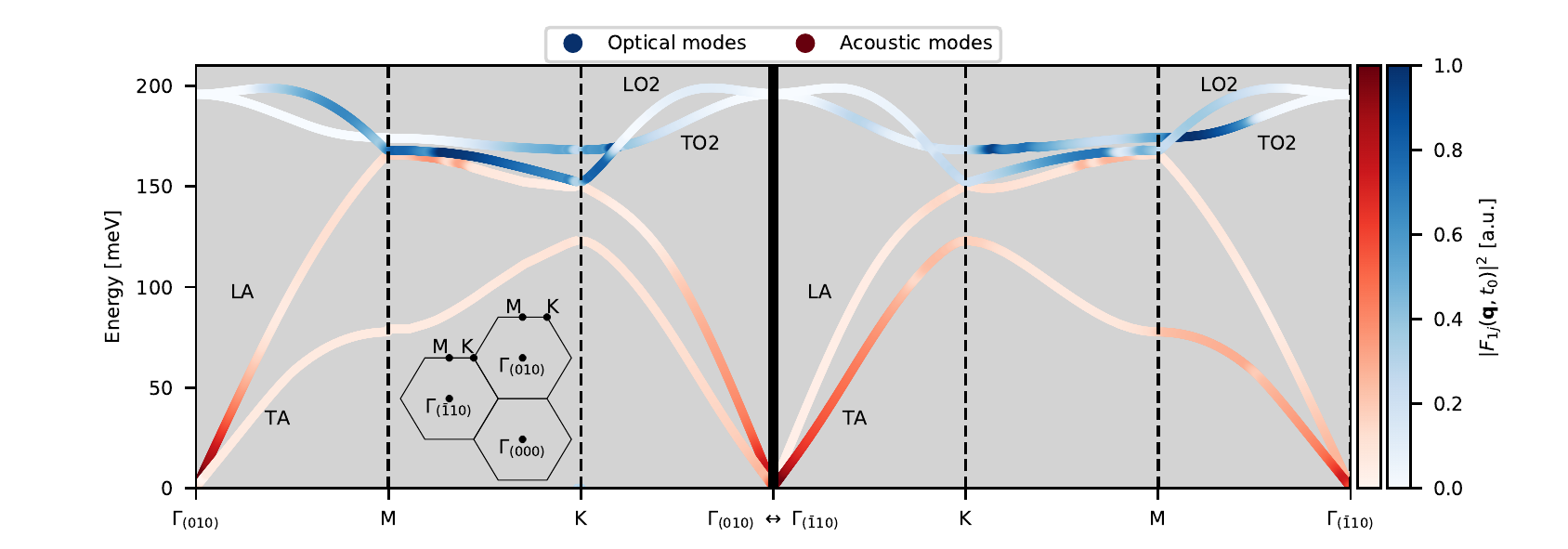}
	\caption{Calculated one-phonon structure factors $|F_{1j}(\vec{q}, t)|^2$ visualized as weighted dispersion curves for selected in-plane modes of graphite, along the high-symmetry lines $\Gamma_{(010)}-M-K-\Gamma_{(010)}$ (left) and $\Gamma_{(\bar{1}10)}-K-M-\Gamma_{(\bar{1}10)}$ (right). The color saturation of dispersion curves is proportional to $|F_{1j}(\vec{q}, t)|^2$ of the associated mode, at \SI{300}{\kelvin} ($t=t_0$). The left and right paths are shown in a diagram on the bottom left. This figure highlights that one-phonon structure factors values are highly variable, and that their values can differ significantly even when comparing neighbouring Brillouin zones. A striking example of this is the relative strengths of the one-phonon structure factors of LA and TA modes near $\Gamma_{(010)}$ and $\Gamma_{(\bar{1}10)}$. At these locations, the ratio of values of $|F_{1j}(\vec{q}, t)|^2$ completely flips, even though the paths are equivalent in the reduced zone scheme.}
	\label{FIG:weighted_dispersion}
\end{figure*}

An alternative view of one-phonon structure factors is presented via weighted dispersion curves, an example of which is shown in Figure \ref{FIG:weighted_dispersion}. This presentation allows easy comparison of the relative weights of the one-phonon structure factors along high-symmetry lines for different phonon branches. 

\begin{figure}
	\centering
	\includegraphics[width=1\columnwidth]{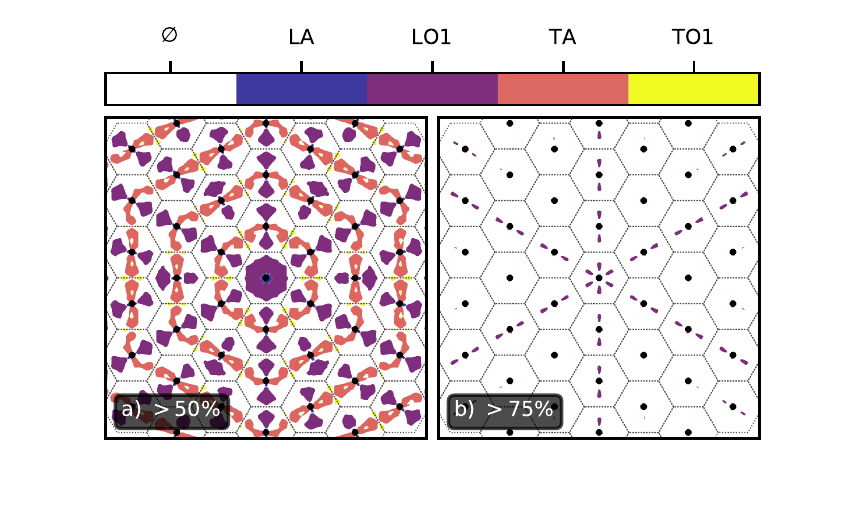}
	\caption{Reciprocal space locations where $|F_{1j}(\vec{q}, t_0)|^2/\omega_{j, \vec{k}}(t_0)$ is dominated by one mode $j$, for $\vec{q}$ vectors equivalent to the detector area shown in Figure \ref{FIG:graphite}. Modes other than those shown here (e.g. TO2) are not dominant anywhere. \textbf{a)} $|F_{1j}(\vec{q}, t_0)|^2/\omega_{j, \vec{k}}(t_0)$ is dominated ($>50\%$) by one mode $j$. White ($\varnothing$) regions account for $63\%$ of the wavevectors, where no phonon mode is dominant. Bragg peaks have been marked by black dots. \textbf{b)}: locations where $|F_{1j}(\vec{q}, t_0)|^2/\omega_{j, \vec{k}}(t_0)$ is dominated ($>75\%$) by one mode $j$. White ($\varnothing$) regions account for $99\%$ of the wavevectors where no mode reaches the $75\%$ threshold.}
	\label{FIG:UEDS_majority}
\end{figure}

A cursory inspection of the weighted dispersion curves in Figure \ref{FIG:weighted_dispersion} suggest that there are regions in the Brillouin zone where diffuse intensity is strongly biased towards a single mode (strong scattering selection rule) based on the relative intensities of one-phonon structure factors. Careful analysis reveals that there are very few wavevectors $\vec{q}$ for which a particular phonon mode's one-phonon structure factor is strongly dominant. Figure \ref{FIG:UEDS_majority} presents a comparison of the relative intensities of one-phonon structure factors weighted by phonon frequency, which is indicative of mode population as per Equation \eqref{EQ:diffuse}. Only $37\%$ of wavevectors visible in measurements shown in Figure \ref{FIG:graphite} have a mode that contributes over $50\%$ of the quantity $\sum_j |F_{1j}(\vec{q})|^2/\omega_{j,\vec{k}}$; only $1\%$ of wavevectors have a phonon mode that contributes more than $75\%$.
The results of Figure \ref{FIG:UEDS_majority} show that quantitative answers regarding phonon dynamics from UEDS measurements cannot generally be obtained by inspection; at almost any wavevector $\vec{q}$, at least two phonon modes contribute significantly to the transient diffuse scattering intensity. Therefore, a more robust procedure, presented in the next section, must be employed to extract wavevector- anndmode-dependent phonon populations from UEDS intensities.
\subsection{Population dynamics across the Brillouin zone}
Transient electron diffuse intensity has been used elsewhere~\cite{Chase2016, Waldecker2017, Stern2018, Konstantinova2018} as an approximation to the population dynamics of particular modes. However, one can extract the transient wavevector-dependent phonon population dynamics $\left\{ \Delta n_{j,\vec{k}}(t) \right\}$ by combining the measurements of $\Delta I(\vec{q},t)$ with the calculations of one-phonon structure factors and associated quantities presented above. 

For many materials (including graphite), the temperature dependence (and hence time dependence) of the phonon mode vibrational frequencies is negligible, because such dependence is proportional to anharmonic couplings between branches~\cite{Calizo2007, Judek2015}; hence, $\omega_{j, \vec{k}}(t) \equiv \omega_{j, \vec{k}}(t_0)$. Moreover, as is discussed in Section \ref{SEC:debye_waller}, the temperature dependence (and hence time dependence) of the Debye-Waller factors --- and therefore the one-phonon structure factors --- has a much smaller magnitude that the variations due to other terms in Equation \eqref{EQ:diffuse}. Therefore, we have $|F_{1j}(\vec{q}, t)|^2 \equiv |F_{1j}(\vec{q}, t_0)|^2$ for all times. In this case, transient scattering intensity at the detector, $\Delta I(\vec{q},t)$, can be expressed as follows:
\begin{equation}
\frac{\Delta I(\vec{q},t)}{N_c I_e} = \sum_j \frac{\Delta n_{j,\vec{k}}(t)}{\omega_{j,\vec{k}}(t_0)} |F_{1j}(\vec{q}, t_0)|^2
\label{EQ:decomposition}
\end{equation}
for $\vec{k}$ away from $\vec{\Gamma}$, where there might be interference with elastic scattering signals.

At every reduced wavevector $\vec{k}$ and time-delay $t$, there are $N$ different values $\left\{ \Delta n_{j,\vec{k}}(t) \right\}$ that must be determined --- one for each phonon mode. Since the one-phonon structure factors $|F_{1j}(\vec{q}, t_0)|^2$ vary over the total wavevector $\vec{q}$, the transient diffuse intensity for at least $N$ Brillouin zones must be considered so that Equation \eqref{EQ:decomposition} can be solved numerically. A linear system of equations must be solved at every reduced wavevector $\vec{k}$. 

Let $\{ \vec{H}_1, ..., \vec{H}_M \mid M \geq N \}$ be the chosen reflections from which to build the system of equations. Then, the transient phonon population of mode $j$ at every $\vec{k}$ and time $t$, $\Delta n_{j,\vec{k}}(t)$, solves the linear system:
\begin{equation}
	\vec{I}_{\vec{k}}(t) = \vec{F}_{\vec{k}} ~ \vec{n}_{\vec{k}}(t)
	\label{EQ:decomp}
\end{equation}
where
\begin{align}
 \vec{I}_{\vec{k}}(t) &= 
 	\frac{1}{N_c I_e}
	\begin{bmatrix}
    	\Delta I(\vec{k} + \vec{H}_1, t) & \dots & \Delta I(\vec{k} + \vec{H}_M, t)
	\end{bmatrix}^T \\
\vec{n}_{\vec{k}}(t) &=
	\begin{bmatrix}
    	\Delta n_{1, \vec{k}}(t)/\omega_{1,\vec{k}}(t_0) & \dots & \Delta n_{N, \vec{k}}(t)/\omega_{N,\vec{k}}(t_0)
	\end{bmatrix}^T \\
\vec{F}_{\vec{k}} &= \nonumber \\
	&\begin{bmatrix}
		|F_{11}(\vec{k} + \vec{H}_1, t_0)|^2 & \dots  & |F_{1N}(\vec{k} + \vec{H}_1,t_0)|^2 \\
		\vdots		   			 			 & \ddots & \vdots					            \\
		|F_{11}(\vec{k} + \vec{H}_M, t_0)|^2 & \dots  & |F_{1N}(\vec{k} + \vec{H}_M,t_0)|^2
	\end{bmatrix}
\end{align}
for $\vec{k}$ vectors away from the $\vec{\Gamma}$ point. These linear systems of equations can be solved numerically, provided enough experimental data ($M \geq N$).

The choice to solve for $\vec{n}_{\vec{k}}(t)$, rather than for the change in population, is related to the degree of confidence that should be placed in the calculation of phonon polarization vectors and frequencies. The phonon polarization vectors are mostly affected by the symmetries of the crystal. On the other hand, phonon vibrational frequencies might be influenced by non-equilibrium carrier distributions. Solving for the ratio of populations to frequencies, rather than populations, is more robust against the uncertainty in the modelling, because the one-phonon structure factors only take into account the polarization vectors. 

We also note that the procedure presented above can be easily extended to (equilibrium) thermal diffuse scattering measurements, where the phonon populations are known at constant temperature, but the phonon vibrational spectrum is unknown. Therefore, using pre-photoexcitation data of a time-resolved experiment, one could infer the phonon vibrational frequencies, which are then used to determine the change in populations using the measurements after photoexcitation. This scheme only relies on the determination of phonon polarization vectors.
\subsubsection*{Wavevector-dependent phonon population dynamics in graphite}
\begin{figure*}
	\centering
	\includegraphics[width=1\textwidth]{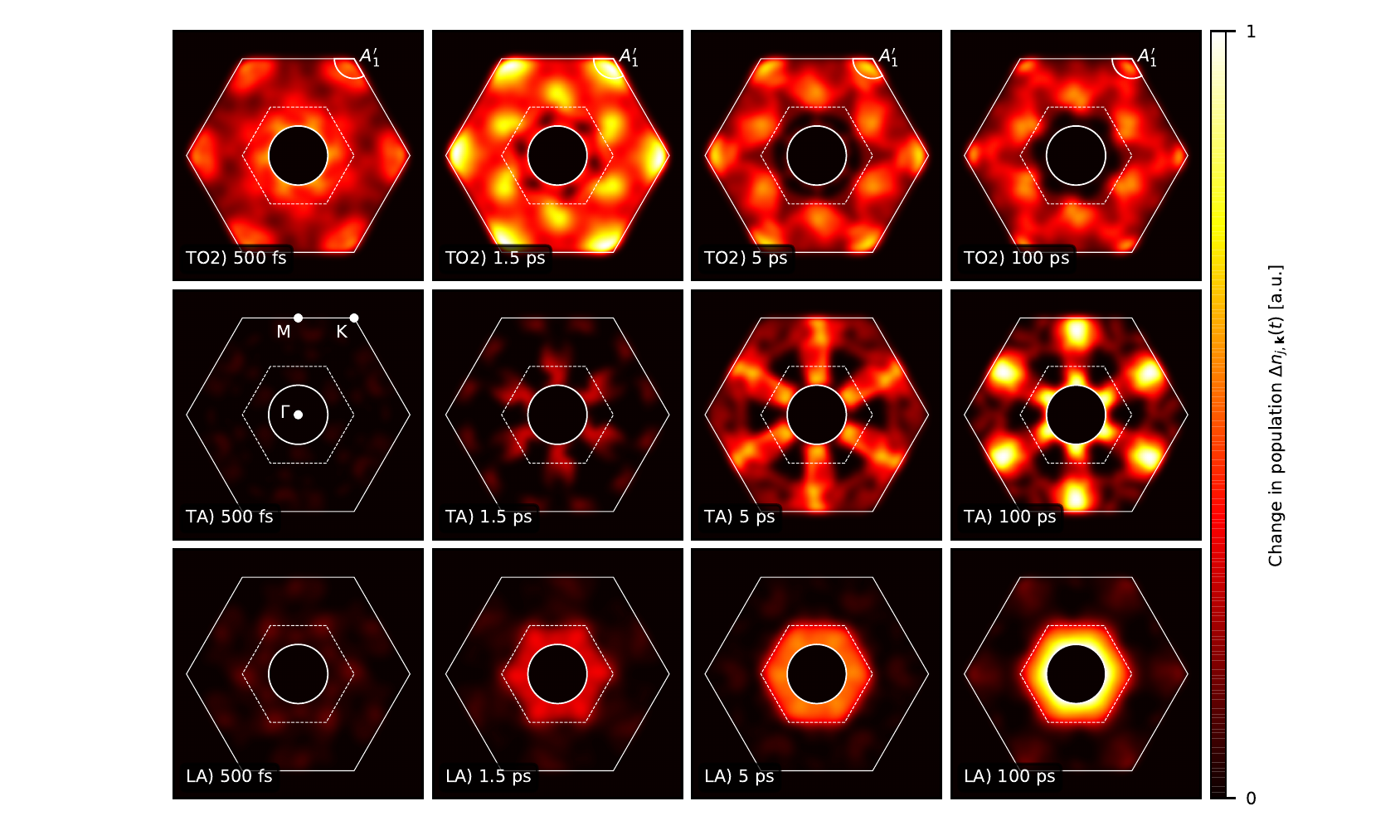}
	\caption{Experimental change in transient population, across the Brillouin zone, of relevant in-plane modes of graphite following photoexcitation. The decomposition of transient diffuse intensity change (Figure \ref{FIG:graphite}) yields stable solutions for $|\vec{k}| > \SI{0.45}{\per\angstrom}$. The bounds of the solutions domain are highlighted by a white circle at $|\vec{k}| = \SI{0.45}{\per\angstrom}$ and by a white hexagon at the Brillouin zone edge. The Brillouin zone midpoints are shown with a dashed hexagon. Uncertainty on the population values is approximately $4\times 10^{-5}$, calculated via the covariance matrix of the matrix solution. The differential $A_1'$ phonon population is highlighted by a circular arc centered at $\vec{K}$ (radius of $\SI{0.3}{\per\angstrom}$) in the TO2 phonon population images; its dynamics are discussed in Section \ref{SEC:a1prime}.}
	\label{FIG:populations}
\end{figure*}

We applied this general formalism for wavevector-dependent phonon population decomposition to the transient diffuse intensity patterns of photoexcited graphite shown in Figure \ref{FIG:graphite}. Since the diffraction patterns have been symmetrized, one would expect that using intensity data for reflections related by symmetry would be redundant. However, better results were achieved by using the entire area of the detector. We expect this is due to minute misalignment of the diffraction patterns and uncertainty in detector position which are averaged out when using all available data. The Brillouin zones associated with all in-plane Bragg reflections $\vec{H}$ such that $|\vec{H}| \leq \SI{12}{\per\angstrom}$ were used, for a total of fourty-four Brillouin zones ($M=44$), many more than the minimum required for the eight in-plane phonon modes of graphite ($N=8$). The physical constraint that $\Delta n_{j,\vec{k}}(t) > 0 ~ \forall t$ was applied~\footnote{The positivity constraint $\Delta n_{j,\vec{k}}(t) > 0 ~ \forall t$ means that the phonon population of a branch cannot drop below its equilibrium level. While not necessary, it leads to more reliable solutions.} via the use of a non-negative approximate matrix inversion approach~\cite{NNLS} to solve Equation \eqref{EQ:decomp} at every reduced wavevector $\vec{k}$ and time-delay $t$. Stable solutions were found for reciprocal space points where $|\vec{k}| > \SI{0.45}{\per\angstrom}$, where there is no interference between elastic $I_0(\vec{q},t)$ and diffuse $I_1(\vec{q},t)$ signals. Figure \ref{FIG:populations} presents the direct decomposition of diffuse intensity into wavevector-dependent, transient phonon population changes $\left\{ \Delta n_{j,\vec{k}}(t) \right\}$, for a few in-plane modes that are particularly relevant to graphite. The discussion of physical processes that explain the wavevector-dependent transient phonon populations follows.

In graphite, optical excitation creates a nonthermal phonon distribution, increasing population primarily in two strongly-coupled optical phonons (SCOP): $A_1'$, located near the $\vec{K}$ point, and $E_{2g}$, at $\vec{\Gamma}$~\cite{Kampfrath2005}. Dynamics measured at the earliest time scales ($<\SI{5}{\pico\second}$) are discussed qualitatively in a previous publication \cite{Stern2018}. Using the measured population dynamics of Figure \ref{FIG:populations}, we can track the transfer of energy across the Brillouin zone quantitatively. \\ By conservation of both momentum and energy, the anharmonic decay of the two SCOP transfers population into mid-Brillouin zone acoustic modes. The early time-points presented in Figure \ref{FIG:populations} confirm that quickly after photoexcitation ($<\SI{500}{\femto\second}$), the transverse optical mode TO2 is strongly populated at $\vec{K}$, indicative of the expected strong electron-phonon coupling to the $A_1'$ phonon. The transfer of energy away from the TO2 mode is already well underway at \SI{5}{\pico\second}, associated with an increase in acoustic modes along the $\vec{\Gamma}$ -- $\vec{M}$ line. This is in accordance with the phonon band structure, where the mid-point along the $\vec{\Gamma}$ -- $\vec{M}$ line favors occupancy of the TA mode (Figure \ref{FIG:weighted_dispersion}). This behaviour intensifies from \SI{1.5}{\pico \second} to \SI{100}{\pico\second}.

The initial increase ($<\SI{5}{\pico\second}$) of TA population along the $\vec{\Gamma}$ -- $\vec{M}$ line is in excellent agreement with predicted anharmonic decay probabilities from the $E_{2g}$ phonon~\cite{Bonini2007}. The (small) increase at \SI{500}{\femto\second} in LA population at $\tfrac{1}{2}\vec{K}$ is also in line with calculated decay probabilities from the $E_{2g}$ phonon by anharmonic coupling~\footnote{On the other hand, a monotonic increase in LA population at $\vec{\Gamma}$ is expected from the decay of the other SCOP, $A_1'$, which has a high probability to decay into a pair of LA-LO modes. However, we were not able to measure population changes close enough to $\vec{\Gamma}$.}.

Over longer time scales ($>\SI{25}{\pico\second}$), the TA population has pooled significantly at $\tfrac{1}{3} \vec{M}$ and $\vec{M}$. There are no three phonon anharmonic decay processes that start in a purely transverse mode; the only allowed interband transitions are L $\leftrightarrow$ T + T and L $\leftrightarrow$ L + T, where L (T) represents a longitudinal (transverse) mode~\cite{Lax1981, Khitun2001}; therefore, a build-up of population in the TA mode is expected. At $\tfrac{1}{2}\vec{M}$, computed lifetimes predict that both LA and TA phonons will favor decay processes into out-of-plane phonons (ZA)~\cite{Paulatto2013} that are not visible (have zero one-phonon structure factor) in the [001] zone-axis geometry in which these UEDS experiments were conducted. Similarily, the ZA phonons predominantly decay back into in-plane phonons, implying that the phonon thermalization in the acoustic branches occurs through a mechanism that exchanges in-plane and out-of-plane modes. Additionally, the computed LA and TA anharmonic lifetimes are predicted to significantly drop at $\tfrac{1}{2}\vec{K}$ and $\tfrac{1}{2}\vec{M}$, respectively, due to the activation of Umklapp scattering to the ZA phonons. Our measurements corroborate these predictions, as can be seen by TA and LA population at the mid-Brillouin zone (dashed white hexagon) being relatively lower than average. The confirmation of those predictions fundamentally relies on UEDS' ability to probe the entire Brillouin zone at once.

The robustness of such an analysis must be emphasized. The decomposition of transient diffuse intensity change via Equation \eqref{EQ:decomp} admits no free parameter. Given sufficient data, a single optimal solution exists.
\subsection{Wavevector-dependent electron-phonon coupling}
\label{SEC:a1prime}
The flow of energy between electronic and phononic subsystems is typically crudely modelled using the \emph{two-temperature} model~\cite{Allen1987}. This model assumes that the electronic system and the phononic system can each be associated with temperatures $T_e$ and $T_{ph}$, throughout the dynamics. Effectively this approximation assumes that the internal thermalization dynamics of each system is much more rapid than any processes that couple the two system. It is evident from the earlier description of the UEDS data from graphite that this assumption is (rather generally) quite a poor one; the idea that the phononic subsystem is internally thermalized does not hold on the timescales typically associated with energy flows between the electron and phonon systems following photoexcitation. On these timescales the phonon occupations are generally very far from being thermalized. UEDS allows to move beyond the two-temperature approximation; by leveraging momentum-resolution, mode-dependent electron-phonon and phonon-phonon couplings can be extracted from the transient change in mode populations $\left\{ \Delta n_{j,\vec{k}}(t) \right\}$. Specifically, wavevector-dependent phonon population dynamics determined in the previous section will now be used to determine the electron-phonon and phonon-phonon coupling strength of the $A_1'$ phonon.

The formalism of the two-temperature model can be extended to the \emph{non-thermal lattice model} (NLM) model~\cite{Waldecker2016}, where every phonon branch $j$ has its own molar heat capacity $C_{ph,j}$, and temperature $T_{ph,j}$:
\begin{equation}
C_e(T_e) \frac{\partial T_e}{\partial t} = \sum_i G_{ep, i}(T_e - T_{ph,i}) + f(t-t_0) 
\label{EQ:mode_dep_ttm_electron}
\end{equation}
\begin{multline}
\label{EQ:mode_dep_ttm_phonon}
\Bigg\{ C_{ph,j}(T_{ph,j}) \frac{\partial T_{ph,j}}{\partial t} = \\
	\sum_{i\neq j} G_{ep, i}(T_e - T_{ph,i}) 
				 + G_{pp,ij} (T_{ph,j} - T_{ph,i}) 	\Bigg\}_{j=1}^{N}
\end{multline}
where $f(t-t_0)$ is the laser pulse profile, and $C_e$ and $T_e$ are the electronic heat capacity and electron temperature, respectively~\footnote{The electronic system thermalizes in approximately \SI{100}{\femto\second}~\cite{Stange2015}, and hence after we can consider the electronic system to be well-described by a single temperature $T_e$.}. This model accounts for discrepancies in coupling between the electronic system and certain phonon modes, which occurs for example in graphite --- where some modes are strongly-coupled to the electron system via Kohn anomalies~\cite{Piscanec2004}. 

Observations of transient changes in mode populations are related to mode temperatures $T_{ph,j}(t)$ via the Bose-Einstein distribution:
\begin{equation}
n_{j,\vec{k}}(t) \propto \left[ \exp\left( \frac{\hbar \omega_{j,\vec{k}}}{k_B T_{ph,j}(t)}\right) -1 \right]^{-1}.
\label{EQ:pop_laurent}
\end{equation}
We can decompose the above expression with a Laurent series~\cite{Wunsch2005} to show explicitly that the mode population is proportional to temperature, for appropriately high $T_{ph,j}$:
\begin{equation}
n_{j,\vec{k}}(t) \propto \frac{k_B T_{ph,j}(t)}{\hbar \omega_{j,\vec{k}}} - \frac{1}{2} + O\left( T_{ph,j}^{-1}(t) \right).
\label{EQ:pop_to_temp}
\end{equation} 
Hence, in the case of measurements presented herein, $\Delta n_{j,\vec{k}}(t) \propto \Delta T_{ph,j}(t)$, where the initial temperature is known to be \SI{300}{\kelvin}.

We now use the NLM to extract the couplings to the $A_1'$ mode from population measurements. The differential $A_1'$ phonon population $\Delta n_{A_1'}(t)$ is obtained by integrating over the region of the wavevector-dependent TO2 phonon population, in a circular arc centered at $\vec{K}$ ($|\vec{k} - \vec{K}| \leq \SI{0.3}{\per\angstrom}$). This location is shown in Figure \ref{FIG:populations}. In order to correlate the mode population measurements with the NLM, the heat capacities of the electronic system and every phonon mode must be parametrized. 

\begin{figure}
	\centering
	\includegraphics[width=3.38in]{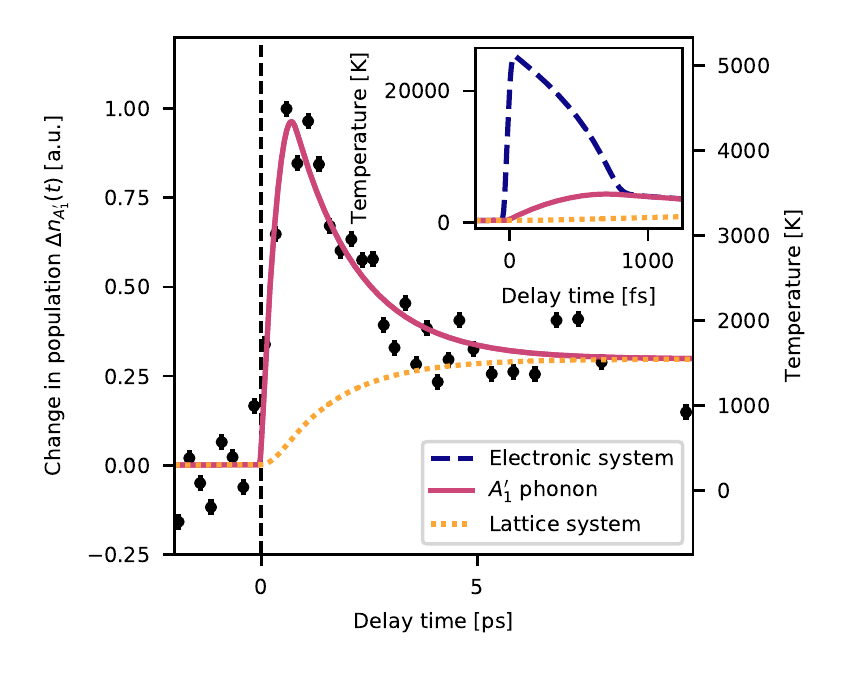}
	\caption{Evolution of the $A_1'$ mode population in graphite after ultrafast photoexcitation. Differential population measurement of $A_1'$, shown in black (circle), is obtained from the integration of the TO2 mode population in a circular arc centered at $\vec{K}$ ($|\vec{k} - \vec{K}| \leq \SI{0.3}{\per\angstrom}$), visible in Figure \ref{FIG:populations}. Error bars are determined from the standard error in the mean of population before photoexcitation ($t<t_0$). The fit to the population change is shown in pink (solid). The effective temperature of the modes in which the $A_1'$ phonon can decay is shown as an orange (dotted) line. \textbf{Inset}: temperature dynamics at early times ($<\SI{1000}{\femto\second}$) show that thermalization between the electronic system (purple, dashed) and the $A_1'$ phonon population (pink, solid) is very fast, indicative of strong electron-phonon coupling. The effective temperature of the modes in which the $A_1'$ phonon can decay is shown as an orange (dotted) line.}
	\label{FIG:a1prime}
\end{figure}

The electronic heat capacity $C_e$ is extracted from experimental work by \citet{Nihira2003}:
\begin{align}
	C_e(T_e) &= 13.8 ~ T_e \\
			 &+ 1.16 \times 10^{-3} ~ T_e^2   \nonumber \\
			 &+ 2.6 \times 10^{-7}  ~ T_e^3.  \nonumber
\end{align}
Over this range of time-delays, thermal expansion (or contraction, in the case of graphite) has not yet occurred~\cite{Chatelain2014a}. No changes in Bragg peak positions --- indicative of lattice parameter changes --- is observed within the experimental range of time-delays $\tau \leq \SI{680}{\pico\second}$. We can therefore calculate the heat capacity of each graphite mode $j$ as the heat capacity at constant volume~\cite{ziman1979principles}:
\begin{multline}
	C_{ph,j}(T_{ph,j}) = \\
	k_B \int_0^{\omega_D} d\omega ~ D(\omega) 
		\left( 
			\frac{\hbar \omega}{k_B T_{ph,j}} 
		\right)^2 
		\frac{e^{\hbar \omega / k_B T_{ph,j}}}{\left( e^{\hbar \omega / k_B T_{ph,j}} - 1\right)^2}
\end{multline}
where $k_B$ is the Boltzmann constant, $\omega_D$ is the Debye frequency, and $D(\omega)$ is the phonon density of states.
Momentum resolution of UEDS allows for a simplification, where a single frequency contributes to the heat capacity in the $A_1'$ mode --- $D(\omega) = \delta(\omega)$. Moreover, we can reduce the number of coupled equations in Equations \eqref{EQ:mode_dep_ttm_electron} and \eqref{EQ:mode_dep_ttm_phonon}. Simultaneous conservation of momentum and energy during the decay of an $A_1'$ phonon can only be satisfied in a few reciprocal space locations. Using first-principles calculations, it is possible to determine the decay probabilities. One such calculation, reported by \citet{Bonini2007}, allows us to define an \emph{effective} heat capacity into which the $A_1'$ population drains, $C_l$~\footnote{This effective heat capacity $C_l$ is composed of 9\% chance to decay into two TA modes, 36\% change to decay into a TA mode and an LA mode, and 55\% chance to decay into either an LA and TA, or LO and LA.}. Therefore, the energy dynamics at $\vec{K}$ can be specified in terms of a system of three equations:
\begin{align}
	C_e(T_e) \frac{\partial T_e}{\partial t}
		&= f(t-t_0) \\
		&- G_{e,A_1'} ~ (T_e - T_{A_1'}) \nonumber \\
		&- G_{e, l}   ~ (T_e - T_l) \nonumber \\
	C_{A_1'}(T_{A_1'}) \frac{\partial T_{A_1'}}{\partial t}
		&= G_{e,A_1'}  ~ (T_e - T_{A_1'}) \\
		&- G_{A_1', l} ~ (T_{A_1'} - T_{l})  \nonumber\\
	C_{l}(T_l) \frac{\partial T_l}{\partial t}
		&= G_{e,l}     ~ (T_e - T_l) \\
		&+ G_{A_1', l} ~ (T_{A_1'} - T_{l}) \nonumber
\end{align}
where $G_{e,A_1'}$, $G_{A_1', l}$, and $G_{e,l}$ are constants. Solving this system of equations gives the temperature evolution of each of the subsystems. The evolution in the $A_1'$ population can be used as a proxy for the mode temperature $T_{A_1'}(t)$; minimizing the difference between observed population dynamics and modelled temperature changes yields the coupling constants $G_{e,A_1'}$, $G_{A_1', l}$, and $G_{e,l}$. The resulting temperature transients are presented in Figure \ref{FIG:a1prime}. The extracted coupling constants have been listed in Table \ref{TAB:coupling}. This model correctly identifies the strong electron-phonon coupling of the $A_1'$ mode, $G_{e,A_1'}$, as compared with the rest of the relevant modes, $G_{e,l}$. 
\begin{table}
	\centering
	\caption{Coupling strength between electronic system, the $A_1'$ phonon, and the lattice system. Uncertainty is derived from fit covariances.}
	\vspace{2mm}
	\begin{tabular}{c c}
		~ & Coupling strength [\si{\watt \per \meter \cubed \per \kelvin}] \\ 
		\hline\\
		$G_{e,A_1'}$  & $(6.8 \pm 0.3) \times 10^{17}$ \\ 
		$G_{A_1', l}$ & $(8.0 \pm 0.5) \times 10^{17}$ \\ 
		$G_{e,l}$     & $(0.0 \pm 6.0) \times 10^{15}$  \\ 
	\end{tabular} 
	\label{TAB:coupling}
\end{table}

From the coupling constant $G_{e,A_1'}$, mode-projected electron-phonon coupling value $\langle g_{e,A_1'}^2 \rangle$ can be determined. In the case of the coupling between the electron system and the $A_1'$ phonon, the heating rate of $G_{e,A_1'} = \SI{6.8 \pm 0.3 e17}{\watt \per \meter \cubed \per \kelvin}$ (Table \ref{TAB:coupling}) corresponds to a mode-projected electron-phonon coupling value of $\langle g_{e,A_1'}^2 \rangle = \SI{0.035 \pm 0.001}{\electronvolt \squared}$ (see appendix \ref{AP:mode_proj_g} for details). These values are in agreement with recent trARPES measurements and simulations~\cite{Johannsen2013, Stange2015, Rohde2018, Na2019}.
\section{Conclusion}
UEDS provides direct access to wavevector-resolved, non-equilibrium phonon populations and is, in this sense, a lattice-dynamical analog of trARPES. A robust and generally-applicable UEDS data reduction method has been described that provides detailed information on transient changes in phonon populations across the entire Brillouin zone that follow photoexcitation in single-crysalline materials. This method takes only the observed UEDS patterns and computed one-phonon structure factors as inputs and can easily be extended with minimal alterations to ultrafast X-ray diffuse scattering. A procedure for computing the required phonon properties using DFPT, and their potential time dependence via the Debye-Waller factors was described in detail. This method was demonstrated for the case of photodoped carriers in the Dirac cones of thin graphite, where the phonon populations were tracked. Finally, the mode dependence of couplings between electron and phonons have been demonstrated at a specific point in the Brillouin zone, where the strongly-coupled optical phonon $A_1'$ is located. Mode-projected electron-phonon coupling value for the $A_1'$ phonon was extracted, using the non-thermal lattice model, and corroborated with numerous other experiments and simulations. 

Direct determination of wavevector-dependent, transient phonon populations holds great promise for the study of phenomena that emerge primarily due to the coupling of electronic and lattice degrees of freedom, and specifically those involving strongly anisotropic interactions. In particular, with sufficient time-resolution, the applicability of the Kramers-Heisenberg-Dirac theory to Raman scattering measurements in graphene/graphite could be explored, via the detection of early-times ($<$\SI{50}{\femto\second}) phonon populations in the strongly-coupled optical phonon $E_{2g}$~\cite{Heller2016}. Another potential extension concerns influence of non-equilibrium carrier distributions on phonon vibrational frequencies. Many systems, charge-density wave materials in particular, exhibit phonon modes that harden or soften at high temperatures and selective electronic excitation which can be used to explore such phenomena in greater depth. 
\section*{Acknowledgements}
L. P. R. de C. thanks H. Seiler for illuminating discussions regarding the role of the Debye-Waller effect on diffuse intensity, and M. X. Na for providing insights regarding the relationship between heat rates and mode-projected electron-phonon coupling constants.

This research was enabled in part by support provided by Calcul Quebec (www.calculquebec.ca) and Compute Canada (www.computecanada.ca)

This work was supported by the Natural Sciences and Engineering Research Council of Canada (NSERC), the Fonds de Recherche du Qu\'{e}bec - Nature et Technologies (FRQNT), the Canada Foundation for Innovation (CFI), and the Canada Research Chairs (CRC) program.
\subsubsection*{Author contribution}
M. S. and B. J. S. conceptualized the work. L. P. R. de C. performed the research. L. P. R. de C. and B.J.S. wrote the manuscript. J.-H. P. performed the DFPT calculations. All authors helped edit the article.
\FloatBarrier
\appendix
\section{Clustering of phonon eigenvalues and eigenvectors into branches}
\label{AP:mode_clustering}

This section describes the clustering of phonon polarization vectors $\left\{ \vec{e}_{j,s,\vec{k}} \right\}$ and frequencies $\left\{ \omega_{j, \vec{k}}\right\}$ into physically-relevant categories, i.e. branches. The general idea behind the procedure is that phonon properties are continuous. The variation of a property should not display any discontinuity along any path in the Brillouin zone.

Let $\vec{P}_{j,\vec{k}}$ be an abstract vector representing the polarization vectors and frequency of branch $j$ at reduced wavevector $\vec{k}$. We represent $\vec{P}_{j,\vec{k}}$ as the following vector:
\begin{equation}
 \vec{P}_{j,\vec{k}} = 
	\begin{bmatrix}
    	\omega_{j, \vec{k}}  &   \vec{e}_{j,s=1,\vec{k}} & \dots & \vec{e}_{j,s=M,\vec{k}}
	\end{bmatrix}^T
\end{equation}
where the index $s$ runs for all $M$ atoms in the unit cell ($M=4$ in the case of graphite). We define the metric between two abstract vectors $\vec{P}_{i,\vec{k}}$ and $\vec{P}_{j,\vec{k}}$ as follows:
\begin{equation}
\lVert \vec{P}_{i,\vec{k}} - \vec{P}_{j,\vec{k}'} \rVert =
	| \omega_{i, \vec{k}} - \omega_{j, \vec{k'}}  |^2 + \sum_s \lVert \vec{e}_{i,s,\vec{k}} - \vec{e}_{j,s,\vec{k}'} \rVert.
\end{equation}
A one-dimensional path $\gamma(\vec{k})$ connecting all $\vec{k}$-points was defined, starting at $\vec{\Gamma}$. At $\vec{\Gamma}$, polarization vectors are associated with a mode based on geometry and oscillation frequency. For example, polarization vectors $\left\{ \vec{e}_{j,s,\vec{k}} \right\}$ parallel to their wavevector $\vec{k}$ for all atoms $s$ is a longitudinal mode; if the associated frequency is $\approx \SI{0}{\tera\hertz}$, this mode can be labelled longitudinal acoustic. Then, polarization vectors and frequencies at any point along the path were assigned to modes that optimized continuity. That is, the assignment of phonon branches $j$ at $\gamma(\vec{k} + \vec{\vec{\Delta}})$, $\vec{P}_{j,\gamma(\vec{k} + \vec{\vec{\Delta}})}$ based on the assignment at $\gamma(\vec{k})$, $\vec{P}_{i,\gamma(\vec{k})}$, minimized the distance 
$\lVert \vec{P}_{i,\gamma(\vec{k})} - \vec{P}_{j,\gamma(\vec{k} + \vec{\vec{\Delta}})} \rVert$. 

The procedure described above, adapted for numerical evaluation, is part of the \texttt{scikit-ued} software package~\cite{RenedeCotret2018}.
\section{Calculation of mode-projected electron-phonon coupling from heating rates}
\label{AP:mode_proj_g}
Consider the coupled equations of the non-thermal lattice model in Equations \eqref{EQ:mode_dep_ttm_electron} and \eqref{EQ:mode_dep_ttm_phonon}. These coupled first-order ordinary differential equations will admit solutions for $T_e(t)$ and $\left\{ T_{ph,j}(t) \right\}$. After photoexcitation ($f(t-t_0) \to 0$), the appropriate summations of those equations yields the following single equation:
\begin{multline}
	\frac{\partial T_e}{\partial t} - \sum_j \frac{\partial T_{ph,j}}{\partial t} =
	\sum_j \Bigg[ \frac{G_{ep,j}}{C_e} ~ (T_e - T_{ph,j})
		   \\-\sum_i \bigg( \frac{G_{ep,i}}{C_{ph,j}} ~ (T_e - T_{ph,i})
		     + \frac{G_{pp,ij}}{C_{ph,j}} ~ (T_{ph,i} - T_{ph,j}) \bigg) \Bigg]
\end{multline}
where the temperature dependence of $C_e$ and $\left\{ C_{ph,j} \right\}$ has been omitted for brevity. In the case of graphite, at early times ($<\SI{5}{\pico\second}$), phonon-phonon coupling $G_{pp,ij}$ is much weaker at the $\vec{K}$-point (Table \ref{TAB:coupling}). Therefore, we may simplify the above equation to a more manageable system:
\begin{align}
	\frac{\partial T_e}{\partial t} - \sum_j \frac{\partial T_{ph,j}}{\partial t}
	&= \sum_j \frac{G_{ep,j}}{C_e} (T_e - T_{ph,j}) \\
	&- \sum_{i,j} \frac{G_{ep,i}}{C_{ph,j}}(T_e - T_{ph,i}). \nonumber
\end{align}
By performing a substitution $\lambda = T_e - \sum_j T_{ph,j}$, the equation above simplifies to a familiar situation:
\begin{equation}
\dot{\lambda}(t) - a(t) \lambda(t) = 0
\label{EQ:Ode}
\end{equation}
where
\begin{equation}
a(t) = \sum_j \left( \frac{G_{ep,j}}{C_e} - \sum_i \frac{G_{ep,i}}{C_{ph,j}}\right).
\end{equation}
The time dependence comes from the time-evolution of the individual temperatures. In the case of phonon temperatures, the phonon population dynamics are directly related to temperature dynamics according to Equation \eqref{EQ:pop_to_temp}. Equation \eqref{EQ:Ode} is a separable equation with solution:
\begin{equation}
\lambda(t) = \exp{\int dt \left[ a(t) \right]}.
\end{equation}
For a slow-varying integrand $a(t) \approx a$, then $a = 1/\tau$, where $\tau$ is a compound variable representing the relaxation of the system. This leads to the following form:
\begin{equation}
\frac{1}{\tau} \approx \sum_j \left( \frac{G_{ep,j}}{C_e} - \sum_i \frac{G_{ep,i}}{C_{ph,j}}\right).
\label{EQ:tconst}
\end{equation}
As a specific example, the above expression reduces nicely in the case of the two-temperature model, where all phonon modes are considered to be thermalized with each other, with isochoric heat capacity $C_{ph}$:
\begin{equation}
\frac{1}{\tau} = G_{ep} \left( \frac{1}{C_e} - \frac{1}{C_{ph}}\right)
\end{equation}
and we see that $\tau$ physically represents the relaxation time of the electronic system into the lattice. Equation \eqref{EQ:tconst} can be thought of as a sum of relaxation times between the electronic subsystem and specific modes $\tau_{e,j}$:
\begin{equation}
\frac{1}{\tau_{e,j}} = \frac{G_{ep,j}}{C_e} - \sum_i \frac{G_{ep,i}}{C_{ph,j}}.
\end{equation}
The final state in relating heating rates to their mode-projected coupling values requires knowledge about density of states. Because the measurements herein consider only in-plane interactions, we use an approximate electronic density of states for graphene close to the Dirac point~\cite{Neto2009}:
\begin{equation}
	D_e(\epsilon) = \frac{2 A}{\pi} \frac{|\epsilon|}{(\hbar~v_F)^2} 
	\label{EQ:graphene_dos}
\end{equation}
where $A$ is the unit cell area and $v_F = \SI{9.06e5}{\meter \per \second}$ is the Fermi velocity~\footnote{Note that factors of $\hbar$ are often ignored, including in the accompanying reference.}. The electronic density of states is related to the mode-projected electron-phonon coupling $\langle g_{ep,j}^2 \rangle$ as follows~\cite{Na2019}:
\begin{equation}
	\frac{\hbar}{\tau_{e,j}} = 2 \pi ~ \langle g_{ep,j}^2 \rangle ~ D_e(\hbar \omega_{\nu} - \hbar \omega_{j, \vec{k}})
\end{equation}
where $\hbar \omega_{\nu}$ corresponds to the optical excitation energy (\SI{1.55}{\electronvolt} or \SI{800}{\nano\meter}).

\end{document}